# Experimental evidence of two-band behavior of MgB$_2$

Ya. G. Ponomarev, S. A. Kuzmichev, M. G. Mikheev, M. V. Sudakova, S. N. Tchesnokov, N. Z. Timergaleev, A. V. Yarigin,
*M. V. Lomonosov Moscow State University, Department of Physics, 119899 Moscow, Russia,*
E. G. Maksimov, S. I. Krasnosvobodtsev and A. V. Varlashkin,
*P. N. Lebedev Physics Institute, RAS, Moscow, Russia,*
M. A. Hein, G. Müller, H. Piel,
*Bergische Universität Wuppertal, Fachbereich Physik, D-42097 Wuppertal, Germany,*
L. G. Sevastyanova, O. V. Kravchenko, K. P. Burdina, B. M. Bulychev,
*M. V. Lomonosov Moscow State University, Department of Chemistry, 119899 Moscow, Russia.*

The break-junction tunneling has been systematically investigated in MgB$_2$. Two types of the break-junction contacts have been exploited on the same samples, which demonstrated tunnel contact like (SIS) and point contact like (SnS) behavior. Both of them have shown the existence of the two distinct energy gaps. We have observed also the peculiarities on the I(V)- characteristics related to Leggett's collective mode assisted tunneling.

The discovery [1] of superconductivity in MgB$_2$ with an exceptionally high transition temperature for such a simple compound (T$_c$ ~ 39 K) has attracted considerable attention. The existence of a significant B isotope effect [2] strongly suggested a well known phonon mediated BCS superconductivity mechanism. On the other hand, theoretical calculations [3-5] as well as experimental observations (see [6] for a review of experiments) have led to the conclusion that MgB$_2$ belongs to the class of multi-band superconductors. It was found [3-5] that the Fermi surface consists of two tubular networks arising from three-dimensional $\pi$ bonding and antibonding bands, and two nearly cylindrical sheets from the two-dimensional $\sigma$ -bands. Usually, these actually four bands are considered as two effective bands. The superconductivity arises from strong electron-phonon coupling in the 2D $\sigma$ -bands. The main source of the superconductivity behavior of the $\pi$ -electrons is their interband coupling to $\sigma$ -electrons. As a result there are two distinct energy gaps. The 2D band shows a large gap, $\Delta_\sigma$ ~ 7-8 meV, whereas the 3D band has a small gap $\Delta_\pi$ ~ 2-3 meV, both closing at the same critical temperature T$_c$ = 39 K.

Multi-band superconductors have been studied intensively since the original theoretical works [7, 8]. One of the most intriguing contributions has been done by Leggett [9]. He has shown that a specific type of collective excitations can exist in two-gap superconductors, which corresponds to small fluctuations of the relative phase of two superconducting condensates. However, these collective excitations have been never observed experimentally in conventional superconductors. The discovery of the two band superconductivity in MgB$_2$ has renewed the interest in Leggett's modes [10-12] and some other phenomena related to the existence of several different superconducting condensate phases [13, 14].

Tunneling spectroscopy is one of the powerful tools to measure the superconducting energy gap. A number of such measurements have been performed on MgB$_2$. A short review of the obtained results is given in paper [15]. There is some evidence of the two-gap behavior in these experiments. However, there is also an ambiguity in the detailed interpretation of tunneling data concerning the existence of two gaps. Some of these measurements show only one gap with a magnitude smaller than the BCS value of $\Delta$ = 1.76 T$_c$. A theoretical investigation of the multi-band model for tunneling in MgB$_2$ junction was done recently in the paper [16]. It was shown that there is a possibility to observe either one or two gaps in the tunneling spectra of MgB$_2$, depending on the tunneling direction, barrier type and impurity concentration.

In the present investigation the current-voltage characteristics (CVC) of more than 150 break-junctions in polycrystalline MgB$_2$ samples have been studied in the temperature range 4.2 K ≤ T ≤ T$_c$. The break-junction technique allows no detailed information about the junction geometry or intrinsic junction properties. The reproducibility of CVC's obtained with this technique is also worse than with some others. However, these shortcomings are partly compensated by a possibility to readjust the contacts easily with a micrometer screw at the helium temperature. Thus the break-junction technique allows changing the junction properties during the measurements, so that the tunnel contact like (SIS) and point contact like (SnS) behavior could be investigated on the same sample. An additional advantage of this technique is the existence of clean cryogenically cleaved surfaces used for the contact formation.

We have used in this investigation three sets of MgB$_2$ polycrystalline samples. The first two sets of MgB$_2$ samples (BG series and BBS series) were prepared by Bulychev, Gentchel (Department of Chemistry, MSU) and Bulychev, Burdina and Sevastyanova (Department of Chemistry, MSU) respectively. The third set of MgB$_2$ samples (KV series) has been prepared by Krasnosvobodtsev and Varlashkin (Physics Institute, RAS). Different techniques of preparation have been used. For BBS and KV series the onset of the resistive transition was T$_{c, onset}$ = 40.5 K with the width ~ 0.3 K. For a BG series the onset of the resistive transition was T$_{c, onset}$ = 39 K with a much wider width ~ 10 K. A local critical temperature T$_c$ in submicron MgB$_2$-break junctions has been determined from the temperature dependence of a superconducting gap Δ(T). For break junctions in BBS- and KV-samples T$_c$ varied in between 35 K and 40.5 K. For junctions in BG-samples a local T$_c$ changed from 22.5 K to 36 K.

Figure 1 represents CVCs of a MgB$_2$ break-junction demonstrating two discussed above types of behavior. One of them corresponds to a typical tunneling junction of the SIS type (curves 1 and 1') with the DC Josephson current and a gap feature at the bias voltage V$_g$ = 2Δ /e. The other one (curves 2 and 2') corresponds to a point contact of the SnS type. The main features of the CVCs of such point contacts comprise an excess current and a subharmonic gap structure (SGS), showing sharp dips of a differential conductance dI/dV at bias voltages:

$$V = \frac{2\Delta}{en} , \qquad \text{with } n = 1, 2... \qquad (1)$$

Usually these SGS are explained by multiple Andreev reflections at the point contact SN-interfaces [17]. This type of behavior of "break-junction" contacts has been observed many times for conventional superconductors [17] and also for high-T$_c$ superconductors [17, 18]. The CVCs of SIS- and SnS- contacts shown on the Fig. 1 demonstrate clearly only a small gap Δ$_\pi$ = 1.8 meV. We would like to emphasize that the values of the gap obtained from the differential conductance dI(V)/dV in the tunneling regime and from the SGS in the point contact regime coincide very well. In the discussion following below we shall consider only the results obtained on the break-junctions giving the same values of superconducting gaps in both regimes.

The results presented in Fig. 2 show the existence of two gaps: the large gap Δ$_\sigma$ = 7.6 ± 0.4 meV and the small gap Δ$_\pi$ = 1.9 ± 0.1 meV. In the point contact regime there is a series of dips for n = 1 and 2 for both gaps. We would like to mention here that the values of the gaps and their temperature dependence can be observed much easier using the point contacts due to a small width of these dips. The temperature dependences of both gaps shown in Fig. 3 were obtained namely from CVCs of point contacts. The temperature dependence of a large gap Δ$_\sigma$(T) is qualitatively close to the BCS type but the ratio $2\Delta_\sigma/kT_c = 5.3$ surpasses the BCS value: 3.52. The temperature dependence of a small gap $\Delta_\pi(T)$ can be different for different junctions. In all cases $\Delta_\pi(T)$ deviates significantly from the BCS type behavior which could be the result of the "intrinsic proximity" effect ("proximity" effect in k-space). However, both gaps ($\Delta_\sigma$ and $\Delta_\pi$) close at one and the same critical temperature $T_c = 34.5$ K.

We have found that the typical values of a characteristic voltage $V_c = I_cR_n$ for the investigated MgB$_2$ Josephson junctions lie within the range $3.0 - 6.0$ mV in good agreement with the theoretical predictions [16]. This result supports indirectly the validity of the two-gap model for MgB$_2$.

We would like now to discuss shortly some unexpected findings of our investigations. For some of MgB$_2$ - break junctions with a local critical temperature $T_c \cong 37 - 40.5$ K we have observed the values of a large gap $\Delta_\sigma$, which exceeded significantly the theoretical estimations [3-5]. Both in tunneling and point-contact regimes we have obtained the values of the large gap of the order $\Delta_\sigma \cong 9 - 11$ meV which leads to the ratio $2\Delta/kT_c \cong 5.6 - 6.3$. The large gap of the same order has been also observed using Andreev spectroscopy by Li et al. [19] and by Takasaki et al. [20] on SIS junctions. It can indicate that a much stronger electron-phonon interaction may exist in the 2D Fermi sheets than was previously anticipated. In any event this problem should be investigated more carefully both theoretically and experimentally.

While investigating SnS contacts with different local $T_c$ (see for example Fig. 3) we have found that only a big 2D gap $\Delta_\sigma$ scaled with $T_c$ (Table 1). At the same time a small 3D gap $\Delta_\pi$ showed no reasonable tendency to change in the range $25 K \leq T_c \leq 40.5$ K.

Another and even more unexpected result is shown in the Fig. 4. We have observed CVCs of stacks of SIS MgB$_2$ junctions (with up to 5 SIS contacts in a stack). The normalized CVCs show clearly the existence of two gaps for all stacks. Moreover, the large gap feature in these CVCs is more and more pronounced with the increasing number of SIS contacts in a stack. This fact allows to measure the temperature dependence of the gap $\Delta_\sigma(T)$ with good accuracy. Until recently, such phenomenon has been observed only in cuprate high-$T_c$ superconductors and it was related to the intrinsic Josephson effect (IJE) between the superconducting blocks of CuO$_2$ – planes in **c**-direction [21-23]. The same results have been obtained also for MgB$_2$ break junctions in the point contact regime. We have registered CVCs of stacks of SnS Andreev contacts ( up to 6 SnS contacts in a stack) with sharp SGS corresponding to $\Delta_\sigma$ and $\Delta_\pi$. Earlier subharmonic gap structures with strong interference pattern have been reported for stacks of SnS contacts in Bi-2201 (intrinsic multiple Andreev reflections effect (IMARE)) [24]. It is very unlikely that the observed effect could be caused by the formation in our break-junctions of a number of stacked, physically equivalent small MgB$_2$ flakes. We believe that this effect is related to the layered structure of MgB$_2$ and to the existence in boron planes of two-dimensional σ - bands which have a weak interband coupling with three-dimensional π -bands. It should be noted that an observation of two-gap structures in the CVCs of stacks of SIS or SnS contacts proves a two-gap superconductivity to be an essentially intrinsic property of MgB$_2$.

As we have mentioned above, multi-band superconductors can possess some special collective excitations due to different values of the order parameter phases in different bands. The existence of such collective modes has been predicted by Leggett [9] and discussed in detail recently in the paper [11] where the expression for the energy $\omega_0$ of this mode has been presented:

$$\omega_0^2 = 4\Delta_1\Delta_2 \frac{\lambda_{12} + \lambda_{21}}{\lambda_{11}\lambda_{22} - \lambda_{12}\lambda_{21}}, \qquad (2)$$

This formula coincides in fact with that of obtained by Leggett [9]. The values $\lambda_{12}$ and $\lambda_{21}$ give the interband coupling between electrons from different bands, $\lambda_{11}$ and $\lambda_{22}$ are the intraband coupling constants and $\Delta_1$, $\Delta_2$ are corresponding energy gaps. Using the values of these parameters, obtained in the first-principle calculations [3, 25], the authors of the paper [11] have estimated the plasma mode energy for MgB$_2$: 6,5 meV $\leq \omega_0 \leq$ 8,9 meV. As it was pointed out in the work [11], to be experimentally observable a Leggett's mode should have the value of $\omega_0$ smaller than twice the smallest gap $2\Delta_1$. Taking into account that $\Delta_1 \cong 2$ meV in MgB$_2$ they concluded that it is unlikely that this mode could be observed. Nevertheless, we have clearly observed the manifestation of this mode with the energy $\omega_0 \cong 4$ meV in the CVCs of MgB$_2$ Josephson junctions at T < T$_c$.

The resonance coupling of the AC Josephson current with some other excitations existing inside the contacts have been observed many times. The theory of such phenomenon for the interaction with electromagnetic waves has been done in the paper [26]. The coupling of the AC Josephson current with the optical phonons has been studied theoretically in [27, 28]. The latter effect was observed experimentally in Bi-2212 break junctions in the phonon frequency range up to 20 THz [29] and in Bi-2212 mesa structures at frequencies up to 6 THz [21, 22, 28]. Shortly speaking, the resonance coupling leads to an enhancement of the DC current flowing throw a contact when a bias voltage V$_{res}$ matches the energy of the corresponding excitations (in this case a "peak-dip" structure of dynamic conductance dI/dV is expected at V$_{res}$). As it was shown in [26], the resonance coupling can exist also between different harmonics of the AC Josephson current and corresponding excitations. The general expression for a voltage V$_{res}$ when the discussed peculiarities become important may be written as:

$$V_{res} = \frac{n}{m} \frac{\omega_0}{2e} , \qquad (3)$$

where $n$ and $m$ are the integer numbers. These peculiarities are more pronounced in the dI/dV-characteristics.

One example of a resonant structure caused by an interaction of the AC Josephson current with the Leggett's mode can be seen in Fig. 1 at the voltage V $\approx$ 2.5 mV. We have observed also for some of break-junction contacts a series of such peculiarities related to the Leggett's mode as it is shown in Fig. 5. We can find features corresponding the following values of $n$ and $m$: 3/2, 1/1, 1/2, 1/3. The value of the energy of Leggett's mode in contacts is of the order $\omega_0 \cong 4$ meV.

As is well known [30] the resonance coupling with excitations can be also observed in SnS Andreev contacts. In particular, when applying external microwave field to the contact one can register several sets of subharmonic gap structures at bias voltages [30]:

$$V_{n,m} = \frac{2\Delta + m\omega_0}{en} , \qquad (4)$$

where $\omega_0$ is a photon energy. In the present investigation we have repeatedly observed in the CVCs of MgB$_2$ SnS-contacts a reproducible SGS of the type (4) without any external microwave field (Fig. 6). In this case the traditional threshold energy $2\Delta$ could be replaced by $(2\Delta + m\omega_0)$ due to a resonant emission of $m$ Leggett's plasmons in the process of multiple Andreev reflections in the SnS-contact. From SGS corresponding to the big gap $\Delta_\sigma$ (Fig. 6) we have derived the excitation energy $\omega_0 \cong 4$ meV which could be also a manifestation of the existence of low-frequency Leggett's plasma mode in a two-gap MgB$_2$ superconductor.

The form of the CVC's peculiarities shown in Fig. 1 and Fig. 5 is very similar to that observed in the phonon-assisted tunneling [27-29]. Peculiarities of the same type can appear also due to an interaction of the AC Josephson current with electromagnetic waves generated inside the tunneling contacts (Fiske resonances) [31]. Nevertheless, we believe that the peculiarities observed in the present investigation are related namely to the Leggett's collective excitations. There are a few reasons for such a conclusion. Firstly, there are no optical phonons with the energy as low as 4 meV in $MgB_2$. Secondly, the effective interaction between the Josephson current and low energy acoustic phonons as well as electromagnetic waves can exist only in the presence of some type of resonator system inside the junction. Then the observed subgap structure will appear at voltages matching the energies of resonator eigenmodes. We have mentioned above that there is no detailed information about the junction geometry or intrinsic junction properties but it is very unlikely that all our break-junctions, demonstrating the discussed subgap structure, possess absolutely identical resonator systems.

The existence of c-axis Josephson plasma resonances typical for superconducting layered cuprates [32, 33] looks also improbable for $MgB_2$ because of a damping role of 3D $\pi$ - bands.

In conclusion we have observed manifestations of the two gap behavior of the $MgB_2$ including the existence of the Leggett's collective mode. We have observed also some unexpected features of the tunneling CVCs. Those are the existence of high values of the energy gap $\Delta_\sigma$ exceeding the theoretical predictions, the intrinsic Josephson effect (IJE) and the intrinsic multiple Andreev reflections effect (IMARE) similar to that observed in high-$T_c$ superconducting cuprates.

This work was made possible by partial financial support from the Scientific Council of the Russian ANFKS state-sponsored R&D program (the 'Delta' Project), the Russian Foundation for Basic Research (Grants No. 02-02-17915, No. 02-02-16658 and No. 02-02-17353) and the RAS complex program, quantum macrophysics.

**References**

1. J. Nagamatsu, N. Nakagawa, T. Muranaka, Y. Zenitani and J. Akimutsu, Nature (London) **410**, 63 (2001).
2. S.L. Bud'ko, G. Lapertot, C. Petrovich et al., Phys. Rev. Lett. **86**, 1877 (2001).
3. A.Y. Liu, I.I. Mazin and J. Kortus, Phys. Rev. Lett. **87**, 087005 (2001).
4. Y. Kong, O.V. Dolgov, O. Jepsen and O.K. Andersen, Phys. Rev. B **64**, 020501-1 (2001).
5. H. J. Choi, D. Roundy, H. Sun et al., Nature **418**, 758 (2002).
6. C. Buzea and T. Yamashita, Supercond. Sci. Technol. **14**, R115 (2001).
7. V.A. Moskalenko, Phys. Met. and Metall. **4**, 503 (1959).
8. H. Suhl, B.T. Matthias, L.R. Walker, Phys. Rev. Lett. **12**, 552 (1959).
9. A. J. Leggett, Prog. Theor. Phys. **36**, 901 (1966).
10. W. Pickett, Nature **418**, 733 (2002).
11. S.G. Sharapov, V.P. Gusynin and H. Becle, arXiv: cond-mat/0205131 v1.
12. D.F. Agterberg, E. Demler and B. Janko, arXiv: cond-mat/0201376 v1.
13. Y. Tanaka, Phys. Rev. Lett. **88**, 017002 (2002).
14. A. Gurevich, V.M. Vinokur, arXiv: cond-mat/02075.
15. H. Schmidt, J.F. Zasadzinski, K.E. Gray and D.G. Hinks, arXiv: cond-mat/0209447 v1 (to be pulished in Physica C).
16. A. Brinkman, A.A. Golubov, H. Rogalla et al., Phys. Rev. B **65**, 180517 (R) (2002).
17. U. Zimmermann, S. Abens, D. Dikin, K. Keck, T. Wolf, Z. Phys. B **101**, 547 (1996).
18. Ya.G. Ponomarev, N.B. Brandt, Chong Soon Khi at al., Phys. Rev. B **52**, 1352 (1995); Ya.G. Ponomarev, E.G. Maksimov, Pis'ma Zh. Eksp. Teor. Fiz. **76**, 455 (2002).
19. Z.-Z. Li, H.-J. Tao, Y. Xuan et al., Phys. Rev. B **66**, 064513 (2002).
20. T. Takasaki, T. Ekino, T. Muranaka, H. Fujii, J. Akimitsu, Physica C **378-381**, 229 (2002).
21. R. Kleiner, P. Müller, Physica C **293**, 156 (1997).
22. A.A. Yurgens, Supercond. Sci. Technol. **13,** R85 (2000).
23. Ya.G. Ponomarev, Chong Soon Khi, Kim Ki Uk et al., Physica C **315**, 85 (1999).
24. Ya.G. Ponomarev, Kim Ki Uk, M.A. Lorentz at al., Inst. Phys. Conf. Ser. No **167**, 241 (2000). H. Schmidt, M.A. Lorenz, G. Muller at al., Abstracts, 6[th] Intern. Conf. M2S-HTSC-VI, Feb. 20-25, 2000, Houston, Texas, 2C2.6, p. 170.
25. A.A. Golubov, J. Kortus, O.V. Dolgov at al., J. Phys. Condens. Matt. **14**, 1353 (2002).
26. J.R. Waldram, A.B. Pippard, J. Clarke, Phil. Trans. Roy. Soc. **268**, 265 (1970).
27. E.G. Maksimov, P.I. Arseev, N.S. Maslova, Solid State Comm. **111**, 391 (1999).
28. K. Schlenga, R. Kleiner, G. Hechtfischer et al., Phys. Rev. B **57**, 14518 (1998).
29. Ya.G. Ponomarev, E.B. Tsokur, M.V. Sudakova et al., Solid State Comm. **111**, 513 (1999).
30. U. Zimmermann, K. Keck, Z. Phys. B **101**, 555 (1996).
31. I. O. Kulik, Pis'ma v JETP **2**, 134 (1965).
32. M. Machida, T. Koyama, M. Tachiki, Phys. Rev. Lett. **83**, 4618 (1999).
33. H. Shibata, arXiv: cond-mat/0204070 v1 3 Apr 2002.

**Table 1.** Local critical temperature $T_c$, superconducting gaps $\Delta_\sigma$ and $\Delta_\pi$ at T = 4.2 K, the ratio $2\Delta_\sigma/kT_c$ and $\Delta_\sigma/\Delta_\pi$ for SnS contacts in different MgB$_2$ polycrystalline samples. In all cases the critical temperature has been estimated from $\Delta_\sigma$ (T) and $\Delta_\pi$ (T)  (see Fig. 3 for example).

| sample | $T_c$, K | $\Delta_\sigma$, meV | $\Delta_\pi$, meV | $2\Delta_\sigma/kT_c$ | $\Delta_\sigma/\Delta_\pi$ |
|--------|------|------|------|------|------|
| MB12 | 26 | 6.2 | 2.0 | 5.5 | 3.1 |
| MB4 | 27 | 6.8 | 2.3 | 5.8 | 3.1 |
| MB6 | 31 | 7.5 | - | 5.6 | - |
| MB1 | 32.5 | 8.2 | 1.6 | 5.9 | 5.1 |
| KR1MSU | 34.5 | 7.9 | 2.2 | 5.3 | 3.6 |
| KRW2 | 36 | 8.8 | 1.6 | 5.7 | 5.5 |
| KRW5 | 37 | 7.9 | 1.8 | 5.0 | 4.4 |
| KR4_T | 40 | 11.0 | 1.8 | 6.4 | 6.1 |

FIGURE CAPTIONS

Fig.1. Small-gap structures ($\Delta_\pi = 1.8$ meV) in the I(V)- and dI/dV-characteristics of a MgB$_2$ break junction in the tunneling regime (SIS contact - 1,1'; dI/dV-peaks can be seen at $V_g = \pm 2\Delta_\pi/e$) and point-contact regime (SnS contact - 2,2'; SGS comprises dI/dV-dips at bias voltages $V_n = \pm2\Delta_\pi/en$ with n = 1 and n = 2) at T = 4.2 K (BBS series).

Fig. 2. Two-gap structures ( $\Delta_\sigma = (7.6 \pm 0.4)$ meV, $\Delta_\pi = (1.9 \pm 0.1)$ meV) in the I(V)- and dI/dV-characteristics of a MgB$_2$ break junction in the tunneling regime (SIS contact - 1,1', dI/dV-peaks present at $V_{g\pi} = \pm 2\Delta_\pi/e$ and $V_{g\sigma} = \pm 2\Delta_\sigma/e$) and point-contact regime (SnS contact, dI/dV-characteristic – 2; two SGS ($Vn_{\pi,\sigma} = \pm2\Delta_{\pi,\sigma}/en_{\pi,\sigma}$) are detectable with dI/dV-dips corresponding to $n_\pi$ (dashed lines) and $n_\sigma$ (dotted lines)) at T = 4.2 K (BBS series).

Fig. 3. Evolution of superconducting gaps $\Delta_\sigma$ and $\Delta_\pi$ with the temperature for two MgB$_2$ SnS contacts ($T_c = 34.5$ K, $2\Delta_\sigma/kT_c = 5.3$, KR series). Solid lines – BCS-type $\Delta(T)$.

Fig. 4. Two-gap structures in the normalized dI/dV-characteristics of SIS MgB$_2$ contacts (T = 4.2 K, $\Delta_\sigma = 8$ meV, $\Delta_\pi = 1.7$ meV): 1, 2 – stacks of 5 SIS contacts, 3 – a stack of 2 SIS contacts, 4 – a single SIS contact.

Fig. 5. A fine structure in the dI/dV-characteristics of four different SIS MgB$_2$ Josephson junctions caused by coupling of the AC Josephson current to a Leggett's plasma mode with the energy $\omega_0 = 4$ meV (T=4.2K). Dotted lines correspond to the bias voltage $V_{res} = n\omega_0/m2e$ with $n/m$ = 3/2, 1/1, 1/2, 1/3.

Fig. 6. A subharmonic $\sigma$-gap structure with "satellites" in the dI(V)/dV-characteristic of an SnS MgB$_2$ contact (samp. KRW4, T = 4.2K, $\Delta_\sigma = 7.5$ meV). This type of structure at bias voltages $V_{n,m} = (2\Delta_\sigma+m\omega_0)/en$ could be caused by a resonant emission of $m$ Leggett's plasmons with the energy $\omega_0 = 4$ meV in the process of multiple Andreev reflections (dotted lines – $m = 1, 2, 3$, dashed lines – $m =0, n = 1,2,3$).

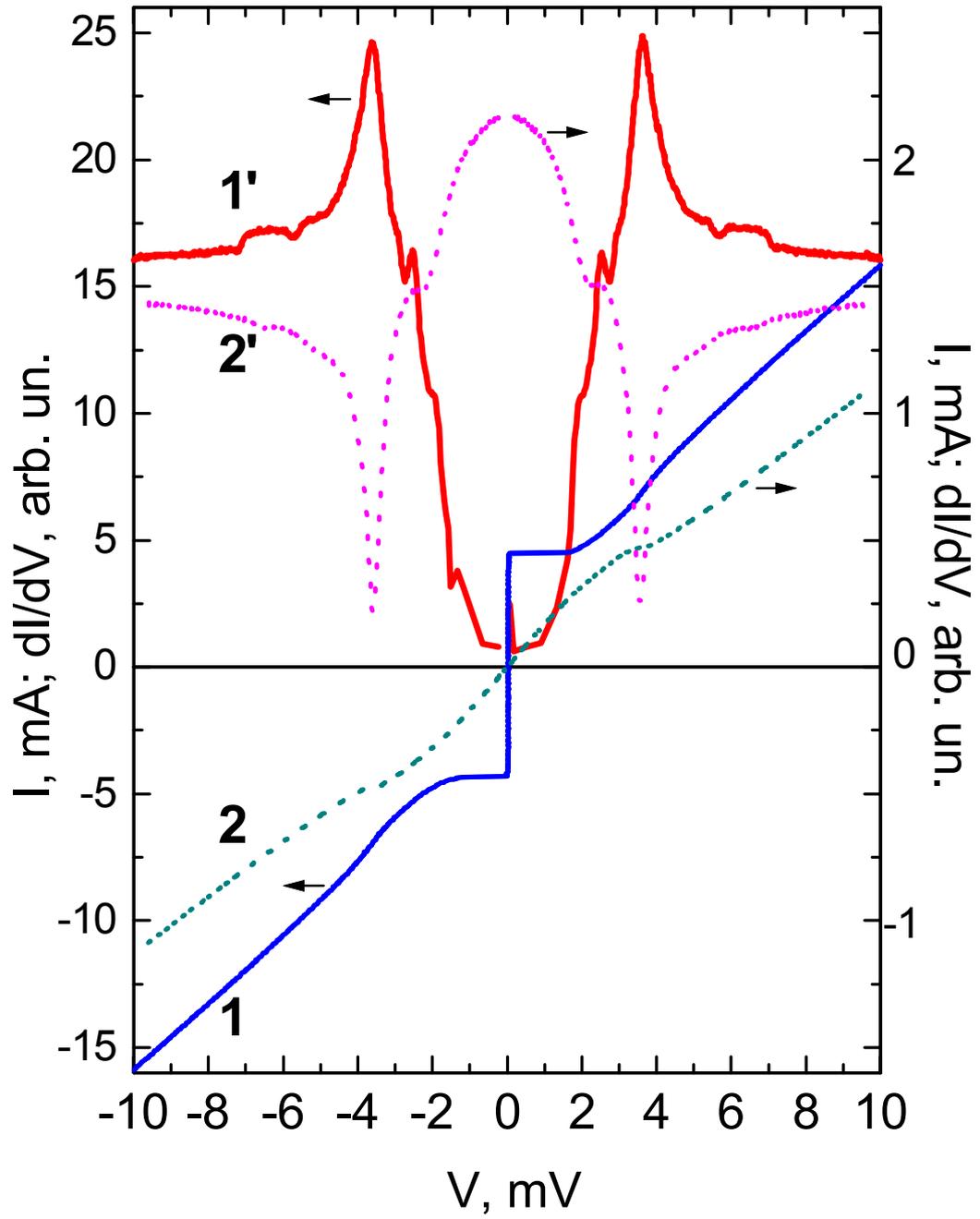

Fig. 1

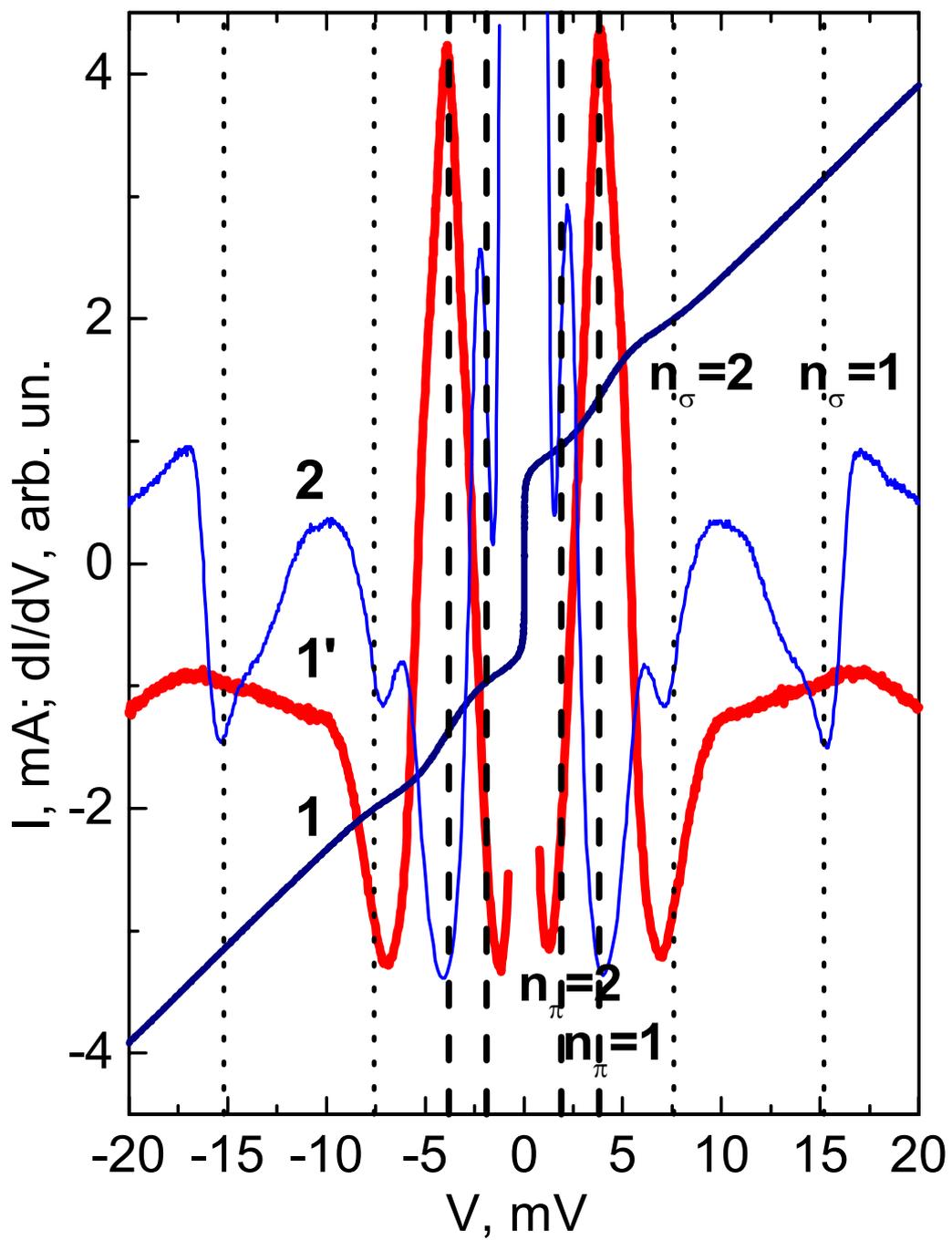

Fig. 2

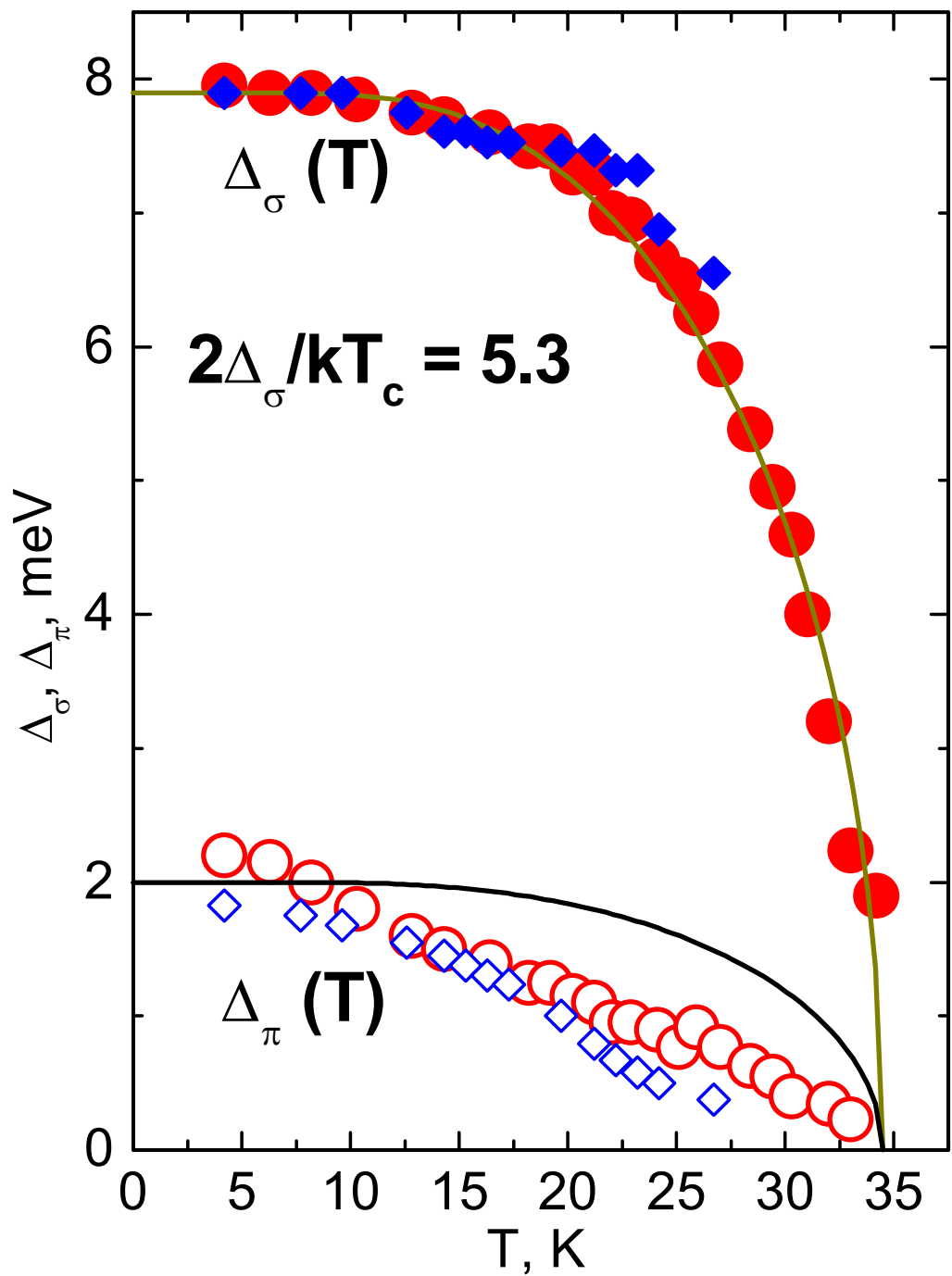

Fig. 3

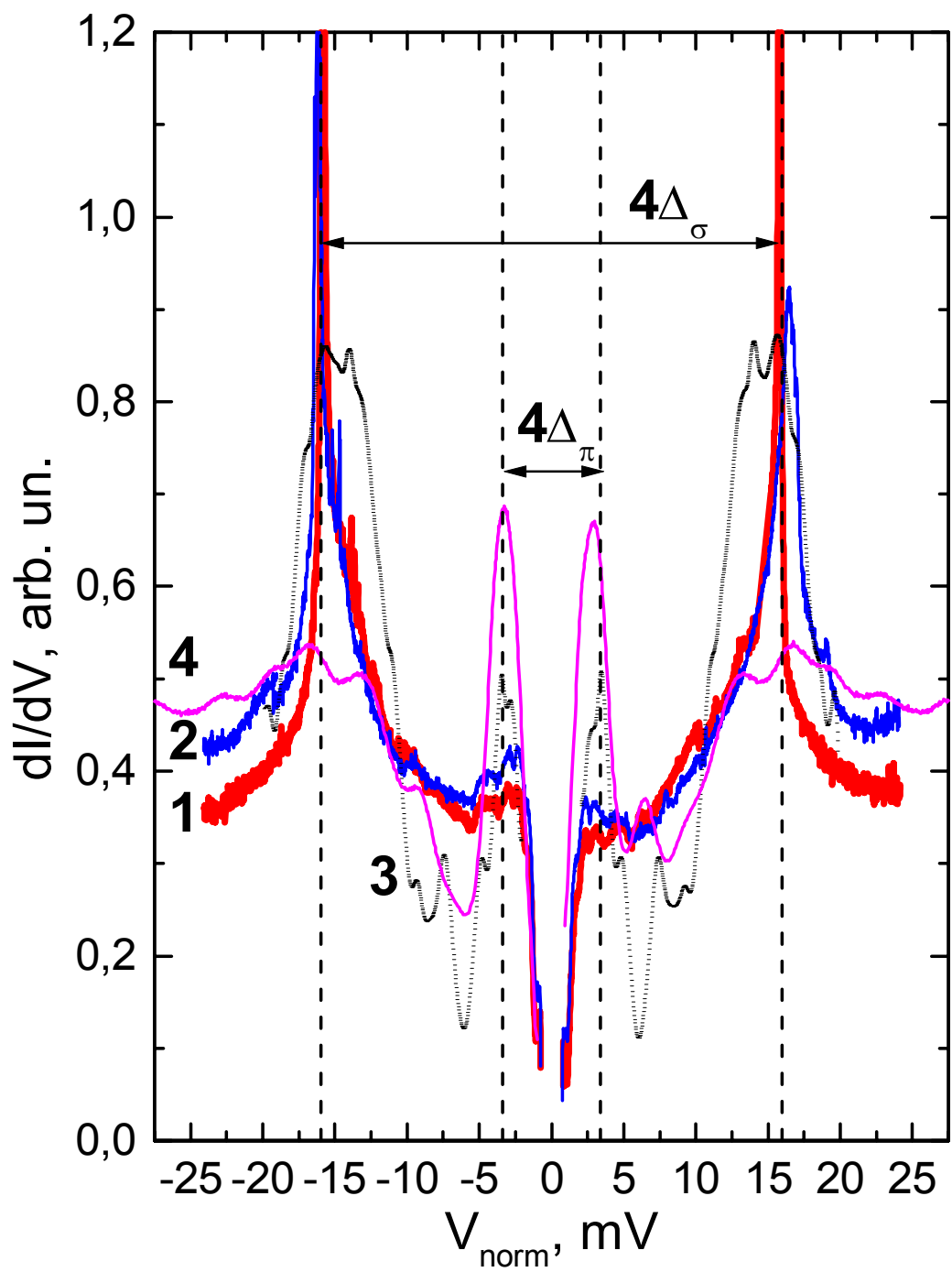

Fig. 4

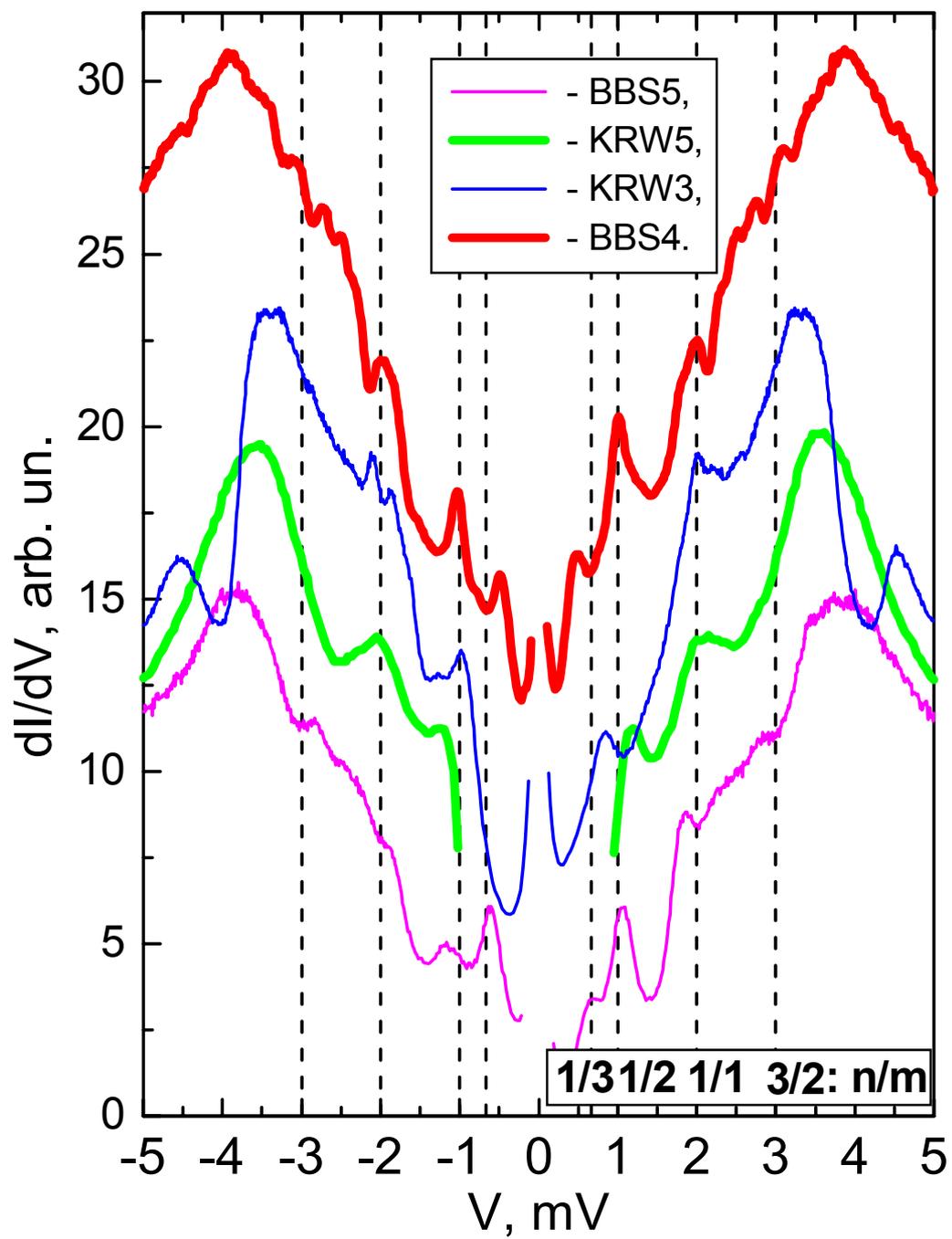

Fig. 5

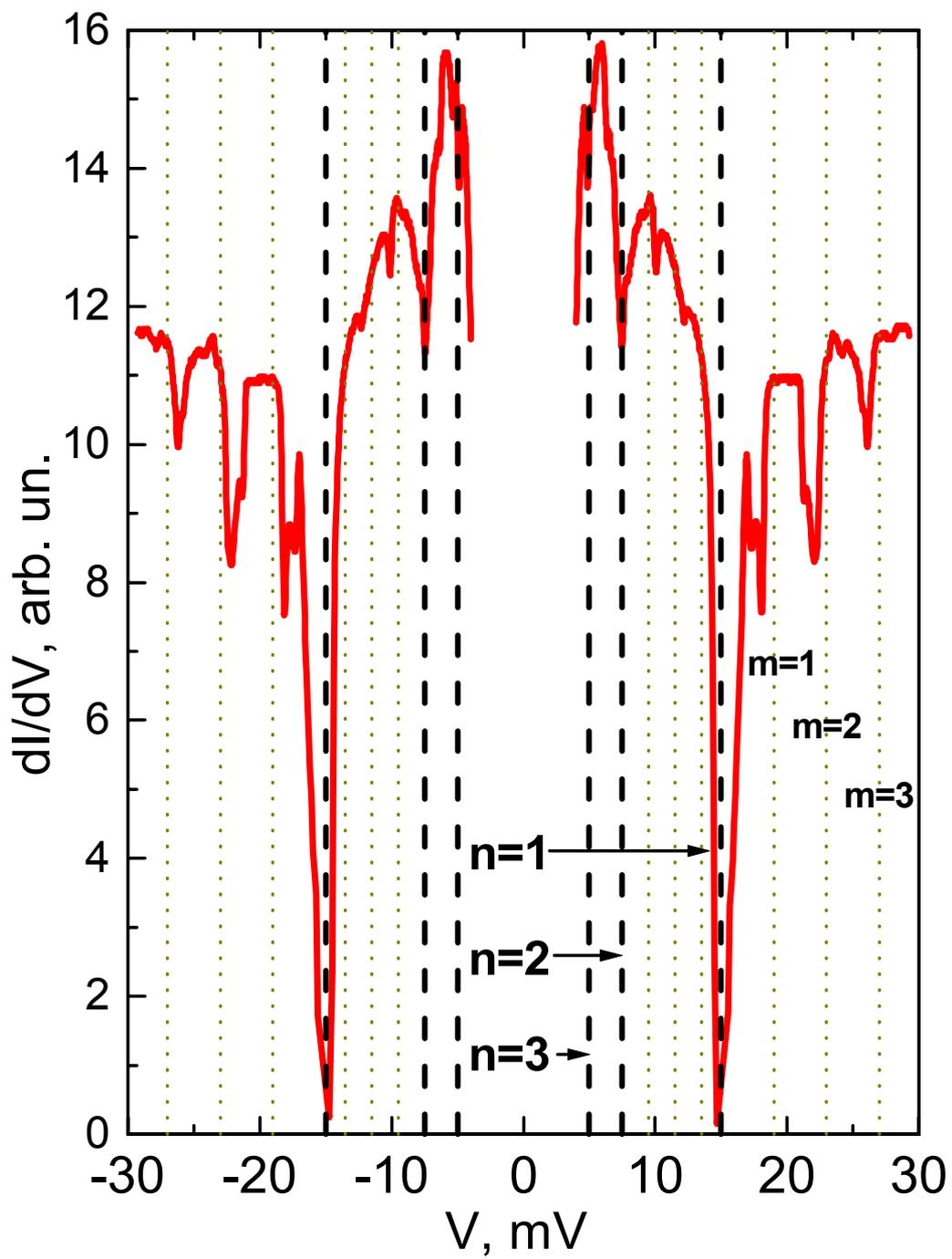

Fig. 6